\documentclass[12pt]{iopart}
\usepackage{times}
\usepackage{graphicx}

\def\PRD{{\em Phys. Rev.} D}

\def\JMP{\em J. Math. Phys.}
\def\bbz{Z\!\!\!Z}
\def\bl{\lambda \kern-6.5pt \lambda}

\def\br{\rho \kern-5.5pt \rho}
\def\bbr{I\!\!R}
\def\bbc{C\kern-6.5pt I}

\def\e{\epsilon}

\def\D{\Delta}

\def\sqr#1#2{{\vcenter{\vbox{\hrule height.#2pt
        \hbox{\vrule width.#2pt height#1pt \kern#1pt
           \vrule width.#2pt}
		\hrule height.#2pt}}}}
\def\square{\mathchoice\sqr68\sqr68\sqr{2.1}3\sqr{1.5}3}		   
\def\xtil{{\lower5pt \hbox{$\buildrel {\rm {\big x}}  \over \sim $ }} }
\def\ptil  {{\lower5pt \hbox{$\buildrel {\big p}  \over \sim $ }} }
\def\qtil {{\lower5pt \hbox{$\buildrel {\big q}  \over \sim $ }} }

\def\simeq {\buildrel {\sim} \over {=} }

\def\pdotp2 { { ( p \cdot {p^'}) }^2 }

\def\bL{\bf L }

\def\bP{\bf P }

\def\bQ{\bf Q }

\def\halb{{1 \over 2}}
\def\vier{{1 \over 4}}
\def\5in2{{5 \over 2}}
\def\3in4{{3 \over 4}}

\def\ddx0{ \matrix{ \partial \cr \overline {\partial x_{0}} \cr } }

\def\ddxi4{ \matrix{ \partial \cr \overline {\partial \xi_{4}} \cr } }

\def\ddu4{ \matrix{ \partial \cr \overline {\partial u_{4}} \cr } }

\def\eq{ ~~=~~}

\def\plus{~~+~~}
\def\minus{~~-~~}

\begin{document}

\title{Representations of Classical Lie Algebras from their Quantum 
Deformations}

\author{P. Moylan}
\address{The Pennsylvania State University}
\address{Abington College}
\address{Abington, Pennsylvania 19001 USA}
\begin{abstract}
 We make use of a well-know deformation of 
the Poincar\'e Lie algebra in $p+q+1$ dimensions ($p+q>0$) to construct 
the Poincar\'e Lie algebra out of the Lie algebras of the 
de Sitter and anti de Sitter groups, the generators of the Poincar\'e Lie 
algebra appearing as certain irrational functions of the generators of the 
de Sitter groups. We have obtained generalizations of 
this ``anti-deformation'' for the $SO(p+2,q)$ and $SO(p+1,q+1)$ cases with 
arbitrary $p$ and $q$. Similar results have been established for $q$ 
deformations $U_q(so(p,q))$ with small $p$ and $q$ values.  Combining known results on 
representations of $U_q(so(p,q))$ (for $q$ both generic and a root of unity) 
with our ``anti-deformation'' formulae, we get representations of classical 
Lie algebras which depend upon the deformation parameter $q$.  Explicit results are given for the simplest 
example (of type $A_1$) i.e. that associated with 
$U_q(so(2,1))$.\footnote{Published in:  Inst. Phys. conf. Ser: No 173: Section 8, 
{\it Proceedings of the 24th International Colloquium on Group Theoretical 
Methods in Physics}, Editors: J-P Gazeau et al., 2003 IOP Publishing Ltd., 
pp. 683-686 }

\end{abstract}

\vskip 1.0truecm
\section{ Introduction. }
We start with a well-known deformation \cite{hm}, \cite{bb} of the Poincar\'e Lie algebra 
in $p+q+1$ dimensions ($p+q>0$), which is defined in terms of the generators ${\bL}_{ij}$ of (pseudo) rotations and the translation generators ${\bP}_i$  by the following:
$${\bL}_{ij} ~~\rightarrow ~~{\bL}_{ij}~~, \eqno(1.1a)$$
$${\bP}_i ~~\rightarrow ~~{\bL}^{\pm}_{p+q+1,i} \eq  {i \over 2~Y} [ {\bQ}_2, ~{\bP}_i] 
~+~{\bP}_i \eqno(1.1b\pm)$$
where ${\bQ}_2~=~ \halb~ \sum_{i,~j~=~0}^{p+q} ~{\bL}_{ij}~{\bL}^{ji}$ is the second order Casimir operator of $SO_0(p+1,q)$, and $Y$ satisfies 
$Y^2 ~=~ \pm ~ \sum_{i,~j~=~0}^{p+q} {\bP}_i ~{\bP}^i$.  ($[~~,~~]$ denotes 
commutator.) Choice of the plus sign in this equation for $Y^2$ 
leads to the Lie algebra of 
$SO_{0}(p+2,q)$ and the minus sign gives the commutation relations of $SO_0(p+1,q+1)$.  Now eqns. (1b$\pm$) may be considered 
as algebraic 
equations for the translation generators  ${\bP}_i $ of the Poincar\'e 
group, and we may attempt to solve these equations for the ${\bP}_i $. The 
solution to this problem for $p=0$, $q=3$ and for the 
choice of 
eqn. (1.b$-$) has been 
given by us in \cite{hm}. The 
general solution 
for the case of eqn. (1.b$+$) ($p=0$, $q=3$) has been presented in 
\cite{m}.  We have also obtained a generalization 
of this ``anti-deformation'' to higher dimensions i.e. we have been able to 
solve eqns.( 1.b$\pm$) for the ${\bP}_i $ \cite{m}, but only by working in 
a particular class of irreducible 
representations, namely that which occurs in the 
decomposition of the left regular representation of $SO_0(p,q)$ groups 
on real hyperbolic spaces \cite{s}. The proof of commutativity of the 
Poincar\'e translation generators for these higher dimensional cases makes 
use of an integral transform \cite{s},  which intertwines certain representations of 
$SO_0(p,q)$ 
induced from the maximal parabolic subgroup with representations which are  
restrictions of the $SO_0(p,q)$ left regular representation 
on eigenspaces 
of the Laplace-Beltrami operator on the hyperbolic space.  

Here we report on some analogous findings for q-deformations of $so(p+1,q+1)$ 
algebras in 
lowest dimensions i.e. for $p+q+1 = 2, 3$ and $4$ \cite{m1} \cite{m2}.  In particular, in the 
$p=1$, $q=0$ case, 
we start with 
the Euclidean group in two dimensions ${\cal E}(2)$, with generators 
${ {\bf L}}_{12}$ (rotation generator) and ${\bf P}_i$ ($i=1,2$) (translation 
generators), and define the following(c.f. \cite{m1}): 
$$ {\tilde {\bf L}}_{3i} =  
\left[{\frac{\left({[-i{\bf L}_{21}]_{{\sqrt q}}}\right)^2}{[2]_{\sqrt q} Y}} 
,{\bf P}_{i} \right] + {\bf P}_{i}~, ~Y := \sqrt{\sum_{i=1}^2 
{\bf P}^{i}{\bf P}^{i} } ~~~~~ ([m]_{q} ~=~  { { q^{m/2} - q^{-m/2} } \over {q^{1/2} - q^{-1/2}}})
. \eqno(1.2)$$
We readily obtain the ``anti-deformation'' by solving eqns. (1.2) 
for the $ {\bf P}_{i}$.  Our results are given below in section 2.

\section{ An Embedding of ${\cal E}(2)$ into a skew 
field extension of $U_{q}(so(2,1))$ . }
\label{typing section}

The q-deformation $U{_q}(so(3, \bbc))$ is defined as the associative algebra over 
$\bbc$ with generators 
$H$, $X^{\pm}$ and relations \cite{m1}, \cite{m2}:
$$ [H, X^{\pm} ] ~=~ \pm 2 X^{\pm} ~~, \eqno(2.1a)$$
$$ [X^{+}, X^{-} ] ~=~  [H]_{q} ~~.  \eqno(2.1b)$$
Let $I$ be the unit element in $U_q(so(3,\bbc))$, then the Casimir element of $U{_q}(so(3, \bbc))$ is 
$$ \Delta_{q} \eq X^{+} X^{-} \plus ([\halb (H \minus I)]_{q})^{2} \minus 
\vier \eq $$
$$ \eq  X^{-} X^{+} \plus ([\halb (H \plus I)]_{q})^{2} \minus 
\vier  \space ~~. \eqno(2.2)$$
The real form $U_{q}(so(2,1))$ of $U{_q}(so(3, \bbc))$ is defined as follows.  
The generators of $U_{q}(so(2,1))$ are given by the following expressions:
$$ {\bL}_{32} \eq \minus {i \over 2} (X^{+} \minus X^{-}) ~~, 
~~ {\bL}_{13} \eq  {1 \over 2} (X^{+} \plus X^{-}) ~~, 
~~ {\bL}_{21} \eq  {{i}\over 2} ~H  ~~.\eqno(2.3)$$
Thus
$$ X^{\pm} \eq {\bL}_{13} \pm i {\bL}_{32}~~.\eqno(2.4)$$
The operators $ ~~ i{\bL}_{12} ~,~~ i{\bL}_{13} ~, ~~ i{\bL}_{32} ~$ are 
preserved under the following antilinear anti-involution $\omega$ of 
$U{_q}(so(3, \bbc))$
$$ \omega(H) = H ~~,  ~~
 \omega(X^{\pm}) = ~ -~ X^{\mp} ~~. \eqno(2.5)$$
For the coproduct on  $U{_q}(so(3, \bbc))$ we take: \cite{gl} 
$$ \D (H) \eq H \otimes I \plus I \otimes H ~, ~~~~~~~~~~~ 
\D (X^{\pm}) \eq X^{\pm} \otimes q^{\frac{H}{4}} \plus  q^{-\frac{H}{4}} 
\otimes X^{\pm} ~. \eqno(2.6)$$

The Lie algebra ${\cal E}(2)$ is the Lie algebra of the Euclidean group, $E(2)$, 
which is the semidirect product of $SO(2)$ with the group of 
translations of the plane, $\bbr^2$.  A basis for the 
Lie algebra ${\cal E}(2)$ consists of the generator of rotations ${
{\bf L}}_{12}$ 
and two commuting translation generators ${\bf P}_{i}$ $(i \eq 1,~2)$.  They 
satisfy the following commutation relations:
$$ [{ {\bf L}}_{12}~,~{\bf P}_{2} ] \eq  ~{\bf P}_{1}~~,~~
[{\bf L}_{12}~,~{\bf P}_{1} ] \eq  - ~{\bf P}_{2} ~~, \eqno(2.7a)$$
$$ [{\bf P}_{1}~,~{\bf P}_{2} ] \eq 0 ~~. \eqno(2.7b)$$
It is useful to work with the complexified translations generators, which 
are:
$$ {\bf P}^{\pm} ~=~ -{\bf P}_{1} ~\pm~ i~ {\bf P}_{2} ~.\eqno(2.8)$$
We also define as above 
$$ {H} ~= ~ - ~2~i~{\bf L}_{2 1}~. \eqno(2.9)$$ 
Then using (2.7) we verify that 
$$ [{ H}, {\bf P}^{\pm} ] ~=~ \pm 2 {\bf P}^{\pm}~,~ [{\bf P}^{+},{\bf P}^{-}]
\eq 0~.\eqno(2.10)$$

We now solve eqns. (1.2) for the ${\bf P}_i$, our solution 
expresses the translation generators 
of  ${\cal E}(2)$ as irrational functions of 
$U_q(so(2,1))$. Thus it gives an embedding of ${\cal E}(2)$ into an 
algebraic extension $K^\prime(U_q(so(2,1)))$ of the skew field $K(U_q(so(2,1)))$ \cite{pk}. 
Explicitly the solution is given by: 
$${\bf P}_1 \eq D^{-1} ~\left(    
\{ I \minus { \frac {1} {2Y} } {\frac { [H]_q } {[H]_{\sqrt {q}} } } \}
{\bf L}_{31} \plus {\frac {i [2]_{\sqrt {q}} } {2 Y} } 
[ {\frac {H} {2} } ]_q {\bf L}_{32} \right) ~~, \eqno(2.11a)$$
and 
$${\bf P}_2 \eq D^{-1} ~\left(    
\{ I \minus { \frac {1} {2Y} } {\frac { [H]_q } {[H]_{\sqrt {q}} } } \}
{\bf L}_{32} \minus {\frac {i [2]_{\sqrt {q}} } {2 Y} } 
[ {\frac {H} {2} } ]_q {\bf L}_{31}\right) ~~, \eqno(2.11b)$$
where  
$$ D \eq \minus {\frac {1 }{4 Y^2} } ~\left\{[ H ]_{\sqrt q}^2 \minus 
( {\frac { [H]_q } {[H]_{\sqrt {q}} } } \minus 2 Y)^2 \right\} ~~~. 
\eqno(2.12)$$
Furthermore
$$  Y^2 \eq { \Delta}_q  \plus \vier~I ~~.~~~~ \eqno(2.13)$$
One readily verifies that the ${\bf P}_{i}$ as defined by eqns. (2.11a) 
and (2.11b) satisfy the defining commutation relations for the translation 
generators of  ${\cal E}(2)$, and 
verify that $Y^2~=~{\bf P}^{+}~ {\bf P}^{-} $.

The embedding given by eqns. (1.2) extends to 
a homorphism $\tau$ from $K^\prime (U_q (so(2,1)))$ to 
$K^\prime (U({\cal E}(2)))$ (an algebraic extension of the skew field of 
$U({\cal E}(2))$. ($U({\cal E}(2))$ is the enveloping algebra of 
${\cal E}(2)$.)  In fact, since ${\bf P}_i$ in (2.11a) and (2.11b) commute, it is easy to see 
that  $\tau$ defined as $  \tau(X^\pm) ={\tilde X}^\pm$ and $\tau(H) =
{H}$  is an isomorphism. If we take the standard coproduct on $U({\cal E} (2))$ \cite{d} and call it ${\tilde \D}$, then one 
verifies that $ \tau(\D(X^\pm)) \ne {\tilde \D} (\tau(X^\pm))$ even for $q=1$. However, we can treat the tensor product of representations 
as in \cite{m2} where we gave a description of $U_q(so(4,\bbc))$ 
similar to the above 
description of $U_q (so(3,\bbc))$.  (It is well-known that $U_q(so(4,\bbc))$ is constructed out of two mutually 
commuting  pairs of $U_q (so(3,\bbc))$ \cite{vd}.)  There we introduced two commuting 
pairs of translation operators defined on the tensor product representation 
of 
two representations of  $U_q (so(3,\bbc))$.  They were defined implicitly 
by equations similar to eqns. (1.2), and, as above for  
$U_q (so(3,\bbc))$, we 
were able to solve the equations for these four translation operators.

A few comments about about the higher dimensional $q$ deformed 
cases: the above 
remarks in the previous paragraph, outline the main ideas of our 
generalization to $U_q(so(2,2))$ 
 and $U_q(so(3,1))$.  ($U_q(so(2,2))$ 
 and $U_q(so(3,1))$ are real forms of $U_q(so(4,\bbc))$.)  We have also obtained a description of the Rac 
representation of $U_q(so(3,2))$ \cite{dm} along these lines.  This uses the fact that 
the Rac representation remains irreducible under  
$U_q(so(2,2))$.

\section{Representations}

For $\sigma \in \bbc$ and for any $q \in \bbc$ ($q \ne 0$ and not a root of unity) the following formulae define a representation $d\pi^{\sigma, \e}$ of $U_q(so(3,\bbc))$ \cite{kac}: ($\e = 0$ or $\frac {1}{2}$)
$$d\pi^{\sigma,\e}(H)\vert m> = 2~m \vert m > ~,~
d\pi^{\sigma,\e}(X^\pm)\vert m> =[-\sigma \pm m]_q \vert m \pm 1> ~~. \eqno(3.1)$$ 
For $q^N \ne 1$ ($|q|=1$):  (1) $\sigma = i \rho - {\frac {1} {2}}$ ($\rho \in 
\bbr$) and the representation space ${\cal D}^{i\rho - 1/2}$ is the linear span of the $\vert m>$ (
$m=n+\e$, $n=0,\pm 1, \pm 2, ~.~.~.$), and $d\pi^{\sigma, \e}$ is the (infinitesmally 
unitarizable) principal series of $U_q(so(2,1))$; (2)  $ \sigma = \e~ {\rm mod}(2)$ and $\sigma = \ell$ 
with $\ell < -{\frac {1} {2}}$ and a) the representation space 
$X_+^{-\ell,\e}$ is the 
linear span of the above $\vert m>$ with $m>-\ell$,   b) the representation space 
$X_-^{-\ell,\e}$ is the 
linear span of the $\vert m>$ with $m<\ell$. $d\pi^{\sigma, \e}$ acts 
irreducibly on $X_\pm^{-\ell,\e}$.  These give $q$ deformed discrete series 
of  $U_q(so(2,1))$.  

For $q^M=1$ ($M \in \bbz$, $M>2$), let $q=e^{\frac 
{2\pi i}{m}}$ and set $M=m$ for $m$ odd, and set 
$M= \frac {m} {2}$ for $m$ even.  Define $\sigma = \frac {1}{2} (d-1) - 
\frac {1}{2} M$ ($d= 1,2,...M$) and let $V_d =$ linear span of the 
$\vert s_3>$ ($s_3 = \sigma, \sigma -1, ... \sigma - (d-1)$). The action 
$d\pi^\sigma$ of the basic generators $H$ and $X^\pm$ on $V_d$ is given 
by: \cite{keller} 
$$d\pi^\sigma (H) \vert s_3> = - 2s_3 \vert s_3>~,~ 
d\pi^\sigma (X^\pm ) \vert s_s> =  [-\sigma \pm s_3]_q \vert s_3 \pm 1>~~.
 \eqno(3.2)$$
These finite dimensional highest weight modules are all infinitesmally 
unitary. 
For which of the 
above representations do eqns. (2.11) determine  a 
representation 
of ${\cal E}(2)$ on the given representation space?  
The following theorem provides the answer to this question. 

\noindent
{\bf Theorem:}  For $q^N \ne 1$ we have representations of 
${\cal E}(2$) on  ${\cal D}^{i\rho - 1/2}$ and on $X_\pm^{-\ell,\e}$ 
but the representation of ${\cal E}(2)$ is infinitesmally unitary only 
on ${\cal D}^{i\rho - 1/2}$. For  $q^N = 1$ ($N\in \bbz$, $N > 2$) none of the 
representations $d\pi^\sigma$ lead to representations of ${\cal E}(2)$ on 
$V_d$.

The main ingredient in the proof of the theorem involves determining the 
action of the operator $D$ of eqn. (2.12) on the given representation 
space, and, in particular, deciding whether zero 
lies in the resolvent set of the operator in its given action on the 
representation space.

\end{document}